\def\fr#1#2{{#1\over #2}}                                   
\def\Tr{\,{\rm Tr}\,}             \def\det{ {\rm det} }
\def\Ra{\Rightarrow}              \def\subsub#1{\paragraph{#1}}

\def\gp{{(+)}}           \def\gm{{(-)}}          \def\gn{{(0)}}
\def\one{{(1)}}          \def\two{{(2)}}         
\def\pd{\partial}        \def\con{\omega}        \def\hcon{\hat\omega}

\def\a{\alpha}           \def\b{\beta}           \def\th{\theta}
\def\m{\mu}              \def\n{\nu}             \def\k{\kappa}
\def\G{\Gamma}           \def\g{\gamma}          \def\ve{\varepsilon}
\def\D{\Delta}           \def\d{\delta}          \def\r{\rho}
\def\S{\Sigma}           \def\s{\sigma}          \def\t{\tau}  
\def\O{\Omega}           \def\o{\omega}          \def\vphi{\varphi}   
\def\L{\Lambda}          \def\l{\lambda}         
\def\cL{{\cal L}}        

                      \def\bO{\bar\Omega}
\def\bE{\bar E}          \def\bv{{\bar v}}       \def\bc{{\bar c}\,}
    \def\bH{\bar H}
\def\hA{\hat A}          \def\hB{\hat B}         \def\hD{\hat D}
     \def\hg{\hat g}         \def\ho{\hat\omega}
  \def\hpd{\hat\partial}  \def\he{\hat e}
\def\hV{\hat V}          
\def\tg{{\tilde g}}           \def\tT{\tilde T}
\documentstyle[11pt,eqs]{article}

\renewcommand{\theequation}{\thesection.\arabic{equation}}

\def\cleq{\setcounter{equation}{0}}   \def\nn{\nonumber}                    
\def\be{\begin{equation}}             \def\ee{\end{equation}}
\def\bea{\begin{eqnarray} }           \def\eea{\end{eqnarray} }
\def\ba#1{\begin{array}{#1}}          \def\ea{\end{array}}
\def\lab#1{\label{#1}}                \def\eq#1{(\ref{#1})}                 
\def\bsubeq{\begin{subequations}}     \def\esubeq{\end{subequations}}
\def\bitem{\begin{itemize}}           \def\eitem{\end{itemize}}  
\def\mb#1{\hbox{\boldmath $#1$}}
\begin{document}

\title{\bf 2D induced gravity from \\
           canonically gauged WZNW system} 
\author{M. Blagojevi\'c\thanks{E-mail address: mb@phy.bg.ac.yu}, 
      ~D. S. Popovi\'c\thanks{E-mail address: popovic@phy.bg.ac.yu}
  ~and B. Sazdovi\'c\thanks{E-mail address: sazdovic@phy.bg.ac.yu}\\
       Institute of Physics, 11001 Belgrade, P.O.Box 57, Yugoslavia}       
\date{}
\maketitle
\begin{abstract} 
Starting from the Kac--Moody structure of the WZNW model for $SL(2,R)$  
and using the general canonical formalism,  we formulate a gauge theory
invariant under local $SL(2,R)\times SL(2,R)$  and diffeomorphisms.
This theory represents a gauge extension of the WZNW system, defined by
a difference of two simple WZNW actions. By performing a partial gauge
fixing and integrating out some dynamical variables, we prove that the
resulting effective theory coincides with the induced gravity in 2D.
The geometric properties of the induced gravity are obtained out of the
gauge properties of the WZNW system with the help of the Dirac bracket
formalism.   
\end{abstract}


\section{Introduction} 

The subject of two--dimensional (2D) gravity has two--fold interest:
first, it describes important dynamical aspects of string theory, as 
an effective theory induced by quantum string fluctuations, and second, 
it represents a useful theoretical model for the realistic theory of
gravity in four dimensions. Being closely related to the Weyl anomaly 
in string theory \cite{1}, the induced gravity features a deep analogy
with the usual Wess--Zumino action in gauge theories, and represents
its gravitational analogue \cite{2}. The effective action for 2D
gravity was originally calculated in the conformal gauge, where it has
the form of the Liouville theory \cite{1,3}. Analyzing the dynamical
structure of this theory in the light--cone gauge Polyakov found an
unexpected connection with $SL(2,R)$ current algebra \cite{2}. The
importance of this result has been confirmed by the existence of a
canonical formulation of the theory in terms of gauge independent
variables, the $SL(2,R)$ currents \cite{4,5}.     

Inspired by the above results, Polyakov studied the connection between
the Wess--Zumino--Novikov--Witten (WZNW) model for $SL(2,R)$ and the
induced gravity in the {\it light--cone gauge\/}, trying to understand
how the geometric structure of spacetime can be obtained out of the
chiral $SL(2,R)$ symmetry of the WZNW model \cite{6} (see also
\cite{7}). Similar approach based on the {\it conformal gauge\/} showed
that the related form of 2D induced gravity, the Liouville theory, may
be obtained from the $SL(2,R)$ WZNW model by imposing certain
conformally invariant constraints \cite{8}. A consistent approach to
this reduction procedure has been formulated using a gauge extension of
the original WZNW model, based on two gauge fields \cite{9}.  
 
In the present paper we shall use the general canonical formalism to 
formulate a gauge theory invariant under local $SL(2,R)\times SL(2,R)$
transformations and diffeomorphisms, which represents a gauge extension
of the WZNW system, 
\be
I(g_1,g_2)=I(g_1)-I(g_2) \qquad g_1,g_2\in SL(2,R) \, ,     \lab{1.1}
\ee
defined by a difference of two simple WZNW actions for $SL(2,R)$ group;
then, we shall show, by performing a suitable gauge fixing and
integrating out some dynamical variables, that the resulting effective
theory coincides with the induced gravity in 2D:    
\be
I_G(\phi,g_{\m\n}) =\int d^2\xi\sqrt{-g}\,\bigl[
  \fr{1}{2}g^{\m\n}\pd_\m\phi\pd_\n\phi +
  \fr{1}{2}\a\phi R  + M\bigl(e^{2\phi/\a}-1\bigr)\bigr]\, ,\lab{1.2}
\ee
We are able to demonstrate this connection in a covariant way, fully 
respecting the {\it diffeomorphism invariance\/} of the induced
gravity, generalizing thereby the results of Polyakov and others
\cite{6,7,8,9}.  

We are going to use the general canonical method of constructing
gauge invariant actions \cite{10}. It is based on the fact that the
Lagrangian equations of motions are equivalent to the Hamiltonian
equations derived from the action 
\bsubeq\lab{1.3} 
\be
I(q,\pi,u)=\int d\xi (\pi_i \dot q^i -H_0-u^m G_m) \, ,    \lab{1.3a}
\ee
where $G_m$ are primary constraints, and $H_0$ is the canonical
Hamiltonian. If $G_m$ are first class constraints, satisfying the 
Poisson bracket algebra
\be
\{ G_m,G_n\}=U_{mn}{^r}G_r \, ,\qquad  \{ G_m,H_0\}=V_m{^r}G_r \, ,
                                                           \lab{1.3b}
\ee
\esubeq 
than the canonical action $I(q,\pi,u)$ is invariant under the following
gauge transformations: 
\bea
&&\d F=\ve^m\{F,G_m\} \, , \qquad F=F(q,\pi) \nn \\
&&\d u^m= \dot\ve^m + u^r\ve^s U_{sr}{^m}+\ve^rV_r{^m} \, . \lab{1.4}
\eea

This paper represents not only an extension of the results obtained in
the previous letter \cite{11}, but also a significant simplification
of the basic dynamical structure; it also gives a natural explanation
of the gauge origin of the geometry of spacetime. The Hamiltonian
approach presented here is in complete agreement with the results of
the Lagrangian analysis \cite{12}.  

We begin our exposition in Section 2 by recalling some basic facts
about the WZNW model for $SL(2,R)$. Then, we use the Hamiltonian 
formalism to analyze chiral symmetries of the model by choosing
$\t=\xi^-$ and $\t=\xi^+$ as the time variables, and derive the related
$SL(2,R)$ currents. 
In Section 3 we use these currents to define the energy--momentum
components $T_\pm$ as the first class constraints satisfying two
independent Virasoro algebras, whereupon the application of the general
canonical formalism leads to the covariant extension (with respect to
diffeomorphisms) of the WZNW model.  
In Section 4 we study the problem of gauging the internal
$SL(2,R)\times SL(2,R)$ symmetry of the WZNW theory by doubling the
number of phase space variables. After defining a new set of
currents $I_{\pm a}$, satisfying two independent $SL(2,R)$ algebras
without central charges, we apply the canonical gauge procedure to the
set of first class constraints $G_m=(T_\pm,I_{\pm a})$, and obtain
our basic model --- canonically gauged action of the WZNW system
\eq{1.1}. 
In Section 5 we define a restriction of the theory based on a subset
of first class constraints $G_m$, then we choose a set of gauge fixing
conditions that does not affect the diffeomorphism invariance,
formulate the quantum action using the BRST formalism, and finally
integrate out some variables to obtain an effective theory that 
coincides with the induced gravity \eq{1.2}. 
In Section 6 we use the Dirac brackets to show how geometric properties
of the induced gravity follow from gauge properties of the WZNW
system, and Section 6 is devoted to concluding remarks. 
Geometric properties of the group $SL(2,R)$ and spacetime manifold
$\S$, as well as some other technical details, are presented in the
Appendix.     

\section{Chiral symmetries of the WZNW model for \mb{SL(2,R)} } 
\cleq

Chiral symmetries of the $SL(2,R)$ WZNW model can be naturally analyzed
in the Hamiltonian formalism based on $\t=\xi^-$ or $\xi^+$ as the
evolution parameters \cite{13}. As a result, one finds that these
symmetries are closely related to the Kac--Moody (KM) structure of the
theory, which plays an essential role in the canonical formalism for
constructing gauge invariant theories. 

\subsection{Construction of the action} 

Two--dimensional WZNW model is a field theory in which the basic field
$g$ is a mapping from $\S$ to $G$, $\S$ being a two--dimensional
Riemannian spacetime, and $G$ being a semisimple Lie group. The model
is defined by the action
\be
I(g)=I_0+n\G=\fr{1}{2}\k\int_\S ({}^*v,v)+\fr{1}{3}\k\int_M (v,v^2)\, , 
                                  \qquad v=g^{-1}dg \, ,     \lab{2.1}
\ee
where the first term is the action of the non--linear $\s$--model,
while the second one is the topological Wess--Zumino term, defined over
a three--manifold $M$ whose boundary is the spacetime: $\pd M=\S$.
Here, $n$ is an integer, $\k=n\k_0$, $\k_0$ being a normalization
constant, $v$ is the Maurer--Cartan (Lie algebra valued) one--form, 
${}^*v$ is the dual of $v$, and $(X,Y)=\fr{1}{2}\Tr(XY)$ is
the Cartan--Killing bilinear form (the trace operation is taken in the
adjoint representation of $G$).  With a suitable choice of $\k_0$ the
Wess--Zumino term is well defined modulo a multiple of $2\pi$, which is
irrelevant in the functional integral $I=\int Dg\exp\bigl[iI(g)\bigr]$.

Using the variation $\d g=gu$ and the first structural equation
$dv+v^2=0$, one obtains  the equations of motion in the form
$d({}^*v-v)=0$. In local coordinates $\xi^\pm$ on $\S$ these equations
can be written as $\pd_-(g^{-1}\pd_+ g) =0$, or, equivalently, 
$\pd_+(g\pd_-g^{-1})=0$.

It should be noted that the WZNW model is invariant under chiral
transformations 
\be
g\to g'=\O(\xi^-)g\bO^{-1}(\xi^+) \, ,                      \lab{2.2}
\ee
where $(\O,\bO)$ belongs to $G\times G$.  

If the group elements are parametrized by some local coordinates
$q^\a$, $g=g(q^\a)$, one can use the expansion 
$v=E^at_a=dq^\a E^a{_\a}t_a$, where $t_a$ are the generators of $G$,
and derive the relations  
\bea
&&({}^*v,v)={}^*dq^\a dq^\b\g_{\a\b}\, ,\qquad 
             \g_{\a\b}(q)\equiv E^a{_\a}E^b{_\b}\g_{ab}\, ,\nn \\
&&(v,v^2)=\fr{1}{2}E^aE^bE^cf_{abc}=-6\,d\t   \nn \, . 
\eea
Here, $\g_{ab}$ is the Cartan metric on $G$, 
$f_{abc}=f_{ab}{^e}\g_{ec}$ are totally antisymmetric structure
constants, and the form of the last equation follows from $d(v,v^2)=0$,
using the theorem that any closed form is locally exact. Then, the
WZNW action \eq{2.1} takes the form  
$$
I(q)=\k\int_\S \bigl(\fr{1}{2}{}^*dq^\a dq^\b\g_{\a\b}
                      - dq^\a dq^\b\t_{\a\b} \bigr) \, ,    
$$
where we used the Stokes theorem to transform the second term into an
integral over $\S$, and $\t=dq^\a dq^\b\t_{\a\b}/2$.  
Choosing the Minkowskian structure for spacetime one can introduce
the inertial coordinates $\xi^\m (\m=0,1)$ on $\S$, and write the
action as 
$$
I(q)=\k \int_{\S} d^2\xi \bigl( 
        {\fr 1 2}\eta^{\m\n}\pd_\m q^\a\pd_\n q^\b \g_{\a\b} 
                  - \ve^{\m\n}\pd_\m q^\a\pd_\n q^\b\t_{\a\b} \bigr)\, .
$$

Now, we turn our attention to $G=SL(2,R)$. Starting from the fact that
any element $g$ of $SL(2,R)$ admits the Gauss decomposition, defined 
by equation \eq{A3}, one can introduce the related group coordinates
$q^\a=(x,\vphi,y)$, use the expressions \eq{A5} and \eq{A6} for
$\g_{\a\b}$ and $\t$, respectively, and derive the following local form
of the WZNW action:  
\bea
I&&=\k \int_{\S}d^2\xi\, \bigl[ 
   \fr{1}{2} \eta^{\m\n}\pd_\m\vphi\pd_\n\vphi
   +2(\eta^{\m\n}-\ve^{\m\n})\pd_\m x\pd_\n y e^{-\vphi}\bigr]\, ,\nn\\
 &&=\k \int_{\S}d^2\xi\, \bigl( 
   \pd_+\vphi\pd_-\vphi +4\pd_+x\pd_-y e^{-\vphi} \bigr)\, , \lab{2.3}
\eea
where $\xi^\pm=(\xi^0 \pm \xi^1)/\sqrt{2}$. 

\subsection{Chiral symmetries and KM currents} 

Chiral symmetries \eq{2.2} are not the standard gauge symmetries: 
parameters of the transformations are not arbitrary functions of both
coordinates, but depend only on $\xi^-$ or $\xi^+$. Usually, gauge
symmetries in the Hamiltonian framework are related to the presence 
of first class constraints. However, if we take $\t=\xi^0$ as the time
variable in the action \eq{2.3}, it is easily seen that there are no 
first class constraints in the theory. The solution to this puzzle
lies in the observation that the Hamiltonian definition of gauge
symmetries is based upon a definite choice of time. The absence of 
gauge symmetries for the choice $\t=\xi^0$ does not mean that these
symmetries are absent for any other choice.
Investigations of 2D induced gravity \cite{14,5} and the WZNW model
\cite{13} showed that the correct approach to understanding chiral
symmetries in the Hamiltonian approach is to use the light--cone
coordinate, $\xi^-$ or $\xi^+$, as the time variable. Following the
approach of reference \cite{13}, we shall be able to detect the chiral
symmetry \eq{2.2} of the $SL(2,R)$ WZNW action \eq{2.3} and find out
its close relationship to a set of currents, satisfying an $SL(2,R)$ KM
algebra.  These currents represent basic objects in our approach: they
will enable us to make a covariant and gauge extension of the WZNW
system \eq{1.1}, whereupon one can dynamically reduce the whole
structure by fixing the gauge and integrating out some dynamical
variables, and obtain the induced gravity action \eq{1.2}.  

\subsub{1.} Let us first consider the choice $\t=\xi^-$,
$\s=\xi^+$. The basic Lagrangian dynamical variables in the action
\eq{2.3} are $q^\a=(x,\vphi,y)$. The definition of the corresponding
conjugate momenta $(\pi_x,\pi_\vphi,\pi_y)$ leads to the following
primary constraints: 
\bea
&&-J_{-x}\equiv \pi_x \approx 0 \, ,  \nn \\
&&-J_{-\vphi}\equiv \pi_\vphi - \k\vphi'\approx 0 \, ,  \nn \\
&&-J_{-y} \equiv \pi_y -4\k x'e^{-\vphi} \approx 0 \, , \nn 
\eea
where prime denotes the space ($\s$) derivative.
It is convenient to transform the constraints 
$J_{-\a}=(J_{-x},J_{-\vphi},J_{-y})$ into the tangent space basis 
by writing $J_{-a}=\bE^\a{_a}J_{-\a}$, where $\bE^\a{_a}$ are the
vielbein components on the $SL(2,R)$ manifold (Appendix A):  
\bea
&&J_{-\gp}=\pi_x \, ,\nn \\
&&J_{-\gn}=x\pi_x+(\pi_\vphi -\k\vphi') \, ,\nn \\
&&J_{-\gm}=-x^2\pi_x -2x(\pi_\vphi -\k\vphi') -4\k x'+\pi_y e^\vphi\, .
                                                           \lab{2.4}
\eea
Poisson brackets of the primary constraints define an $SL(2,R)$ KM
algebra with central charge $c_-=-2\k$:   
\be
\{J_{-a},J_{-b}\}=f_{ab}{^c}J_{-c}\d -2\k\g_{ab}\d' \, .   \lab{2.5}
\ee

Since the Lagrangian is linear in velocities, the canonical Hamiltonian
vanishes, and the total Hamiltonian takes the form
$\bH_T=\int d\s u^a J_{-a}$.
The consistency conditions of the primary constraints $J_{-a}$ are
$$
{d\over d\t}J_{-a} =\{ J_{-a},\bH_T\}\approx 2\k u_a'\approx 0\, ,
$$
which implies $u^a(\t,\s)=u^a(\t)$, i.e. $u^a$ is an arbitrary
multiplier depending on $\t=\xi^-$ only. Therefore,  
$$
\bH_T=u^a(\t)j_{-a}\, ,\qquad j_{-a}\equiv \int d\s J_{-a}(\t,\s)\, .
$$
It follows from \eq{2.5} that the constraints $j_{-a}$, the zero modes
of $J_{-a}$, are of the first class: 
$$
\{ j_{-a},j_{-b}\}=f_{ab}{^c}j_{-c}\d \, .
$$ 

The presence of arbitrary multipliers $u^a(\t)$ in $\bH_T$ means that
the theory possesses a specific gauge symmetry, the chiral symmetry,
characterized by parameters $\o^a=\o^a(\t)$. Since there are no
secondary constraints, the symmetry generator takes the simple form:
$G = \o^a(\t) j_{-a}$ \cite{15}. The symmetry transformations of
$q^\a=(x,\vphi,y)$ are given as  
$$
\d q^\a=\{ q^\a, G \} =-\bE^\a{_a} \o^a\, .
$$
The related symmetry transformations of $g(q^\a)$ are:
$$
g\d g^{-1}=t_a\bE^a{_\a}\d q^\a = -\o \quad\Ra\quad \d g=\o g \, ,
$$
where $\o\equiv \o^a t_a$. Since $\o$ is an infinitesimal parameter, we
can write  
$$
g \to g+\d g\approx \O(\xi^-)g\, ,\qquad \O(\xi^-)=e^{\o(\xi^-)} \, ,
$$
which is equivalent to the $\O(\xi^-)$ piece of \eq{2.2}. Thus, $\xi^-$
chiral symmetry is produced by the zero modes of the KM currents
$J_{-a}$.  

The Hamiltonian equations of motion are 
\bea
&&\dot q^\a =\{ q^\a,\bH_T\}= -\bE^\a{_a}u^a(\t) \, ,\nn  \\
&&g{d\over d\t} g^{-1}=t_a\bE^a{_\a}\dot q^\a = -u  
                              \quad\Ra\quad  \dot g=u g \, ,  \nn
\eea
where $u=u^a t_a$. This implies $\pd_+(g\pd_-g^{-1})=0$, in
conformity with the Lagrangian result. 

\subsub{2.} Now, we consider the second choice,
$\t=\xi^+$,~~$\s=-\xi^-$  (the minus sign is adopted in order to
preserve the orientation of the manifold), and find the following
primary constraints:   
\bea
&&-J_{+x}\equiv \pi_x +4\k y'e^{-\vphi} \approx 0 \, ,\nn \\
&&-J_{+\vphi}\equiv \pi_\vphi +\k\vphi'\approx 0\, ,\nn \\
&&-J_{+y}\equiv \pi_y \approx 0\, .  \nn 
\eea
They can be transformed into the tangent space basis with the help of 
$J_{+a}=E^\a{_a}J_{+\a}$:
\bea
&&J_{+\gp}=y^2\pi_y + 2y(\pi_\vphi+\k\vphi')
                    - 4\k y'-\pi_x e^{\vphi}\, ,\nn\\
&&J_{+\gn}=-y\pi_y-(\pi_\vphi+\k\vphi') \, ,\nn\\
&&J_{+\gm}=-\pi_y\, .                                        \lab{2.6}
\eea
The related Poisson bracket algebra has the form of the KM algebra with
central charge $c_+=2\k$: 
\be
\{J_{+a},J_{+b}\}=f_{ab}{^c}J_{+c}\d +2\k\g_{ab}\d' \, .     \lab{2.7}
\ee

The canonical Hamiltonian vanishes, while the total Hamiltonian is 
linear in $J_{+a}$. The rest of the Hamiltonian analysis can be done in
a similar manner, leading to the $\bO(\xi^+)$ piece of the chiral
symmetry \eq{2.2}.    

\section{Covariant extension of the WZNW model} 
\cleq

In the previous analysis we obtained chiral symmetries of the WZNW
model by using $\t=\xi^{\mp}$ as the evolution parameter in the
Hamiltonian approach. These symmetries are generated by the zero modes
of the KM currents $J_{\mp a}$. Now, we return to the usual formulation
with $\t=\xi^0$, and discuss how the KM structure of the WZNW model can
be used to build the covariant extension (with respect to
diffeomorphisms) of the WZNW model \eq{2.3}. 

Using the explicit, canonical expressions for the KM currents, given
by equations \eq{2.4} and \eq{2.6}, we can construct the related
$SL(2,R)$ invariant expressions, 
\bea
&&T_-(q,\pi)={1\over 4\k}\g^{ab}J_{(-)a}J_{(-)b}=
            {1\over 4\k}\bigl[\pi_x\pi_y e^\vphi
             +(\pi_\vphi -\k\vphi')^2 \bigr] -x'\pi_x\, ,\nn\\
&&T_+(q,\pi)=-{1\over 4\k}\g^{ab}J_{(+)a}J_{(+)b}=
            -{1\over 4\k}\bigl[\pi_x\pi_y e^\vphi
              +(\pi_\vphi +\k\vphi')^2 \bigr] -y'\pi_y\, ,  \lab{3.1} 
\eea
representing the components of the energy--momentum tensor (the
Hamiltonian of the action \eq{2.3} for $\t=\xi^0$ is given by
$T_--T_+$). These components satisfy two independent Virasoro algebras:
\be
\{T_\mp(\s_1),T_\mp(\s_2)\}=-[T_\mp(\s_1)+T_\mp(\s_2)]\pd_1\d\, .
                                                            \lab{3.2}
\ee

The above result shows that the KM currents of the WZNW model can be
used to construct the Virasoro algebra, which is equivalent to the
algebra of {\it diffeomorphisms\/} (see, e.g., Ref. \cite{5}). In the
next step we shall use the general canonical formalism, expressed by
equations \eq{1.3}, to construct a {\it covariant theory\/}, in which 
\bsubeq \lab{3.3} 
\be
H_0=0\, ,\qquad   G_m = (T_-, T_+)\, .                     \lab{3.3a}
\ee
This is done by introducing the canonical Lagrangian
\be
\cL(q,\pi,h)=\pi_\a{\dot q}^\a -h^-T_- -h^+T_+ \, .        \lab{3.3b}   
\ee
\esubeq 
To see the usual content of this Lagrangian, one can eliminate the
momentum variables with the help of the equations of motion:  
\bea
&&\pi_x={4\k\over h^- -h^+}e^{-\vphi}\,(\pd_0 +h^+\pd_1)y \, , \nn\\
&&\pi_\vphi\pm\k\vphi'={2\k\over h^- -h^+}(\pd_0+h^\mp\pd_1)
                                                      \vphi\, ,\nn\\
&&\pi_y={4\k\over h^- -h^+}e^{-\vphi}\,(\pd_0 +h^-\pd_1)x \, . \nn
\eea
Then, after introducing new variables $(h^-,h^+)\to \tg^{\m\n}$,   
$$
\tg^{00}={2\over h^- -h^+} \, ,\qquad 
\tg^{01}={h^- +h^+\over h^- -h^+}\, ,\qquad
\tg^{11}={2 h^-h^+\over h^- -h^+} \, ,
$$
with $\det(\tg^{\m\n})=-1$, one obtains 
\be
\cL(q,h) =\k\bigl[\fr{1}{2}\tg^{\m\n}\pd_\m\vphi\pd_\n\vphi
+2(\tg^{\m\n}-\ve^{\m\n})\pd_\m x\pd_\n y e^{-\vphi}\bigr]\, .\lab{3.4}
\ee
It is now natural to identify $\tg^{\m\n}$ with the metric density,
whereupon the above expression is seen to represent the covariant
generalization of the WZNW theory. 

The transformation properties of $\tg^{\m\n}$ are consistent with this
interpretation. Indeed, using the general form of the gauge
transformations \eq{1.4}, where $V_r{^m}=0$ and $U_{sr}{^m}$ is
calculated from the algebra \eq{3.2} (Appendix C), one obtains 
\bsubeq \lab{3.5} 
\be
\d h^\mp=\pd_0\ve^\mp+h^\mp\pd_1\ve^\mp-\ve^\mp\pd_1 h^\mp\, ,\lab{3.5a}
\ee
Then, after introducing new parameters $\ve^\mp=\ve^1-\ve^0 h^\mp$
one finds 
\be
\d\tg^{\m\n}=\tg^{\m\r}\pd_\r\ve^\n+\tg^{\n\r}\pd_\r\ve^\m
             -\pd_\r\bigl(\ve^\r\tg^{\m\n}\bigr) \, ,        \lab{3.5b}
\ee
\esubeq 
which is the diffeomorphism transformation of the metric density
$\tg^{\m\n}=\sqrt{-g}g^{\m\n}$, as follows from \eq{B5}.

Let us note that the Lagrangian \eq{3.4} is also invariant under
conformal rescalings of the metric. Of particular importance for latter
considerations is  the light--cone basis, defined in Appendix B, in
which the Lagrangian \eq{3.4} can be written as  
\be
\cL (q,h) =\k\sqrt{-\hg}\,\bigl[\hpd_+\vphi\hpd_-\vphi
                      +4\hpd_+x\hpd_-y e^{-\vphi} \bigr]\, ,\lab{3.6} 
\ee
where
$$
\hpd_\pm \equiv \he_\pm{^\m}\pd_\m 
         ={\sqrt{2}\over h^- -h^+}(\pd_0 +h^\mp\pd_1) \, ,
\qquad \sqrt{-\hg}=\fr{1}{2}(h^- -h^+) \, .
$$

\section{Gauging \mb{SL(2,R)\times SL(2,R)} and the WZNW system} 
\cleq

In the previous section we obtained the covariant extension of the WZNW
model, using the energy--momentum components $T_\pm$ as the generators
of {\it diffeomorphisms\/}. Since the components $T_\pm$ satisfy the
Virasoro algebras \eq{3.2} without central charges, they are first
class constraints, and one is able to apply directly the general
canonical method for constructing gauge invariant actions, presented in
the introduction.    

Our next task is to consider the possibility of gauging the 
{\it internal\/} $SL(2,R)\times SL(2,R)$ symmetry. One should observe
that the currents $J_{\pm a}$ are not of the first class, since the
related KM algebras have central charges $c_\pm=\pm 2\k$;
therefore, they can not be used as the gauge generators. We wish to find
a set of generators satisfying two independent $SL(2,R)$ algebras without 
central charges. To this end we double the number of dynamical variables,
$$
q\to (q_1,q_2), \qquad \pi\to (\pi_1,\pi_2) \, ,
$$
and introduce two sets of currents,
\be
J^\one_{\pm a}=J_{\pm a}(q_1,\pi_1) \, ,\qquad
J^\two_{\pm a}=J_{\pm a}(q_2,\pi_2)\vert_{\k\to -\k} \, ,    \lab{4.1}
\ee
satisfying two $SL(2,R)$ KM algebras with opposite central charges:
$$
c^\one_\pm =\pm 2\k \, , \qquad c^\two_\pm =\mp 2\k \, .
$$
Now, we introduce new currents,
\be
I_{\pm a}=J_{\pm a}^\one +J_{\pm a}^\two \, ,               \lab{4.2}
\ee
which are easily seen to satisfy two independent $SL(2,R)$ algebras with 
{\it vanishing central charges\/}. The new currents are of the first
class, and can be used to gauge the internal $SL(2,R)\times SL(2,R)$
symmetry. 

In order to include the diffeomorphisms into this procedure, we
introduce the energy--momentum components of two sectors, defined in
terms of $J^\one$ and $J^\two$ as in \eq{3.1},
\be
T_\pm^\one =T_\pm(q_1,\pi_1)\, ,\qquad 
T_\pm^\two =T_\pm(q_2,\pi_2)\vert_{\k\to -\k}\, ,           \lab{4.3}
\ee
which obey the Poisson bracket algebra \eq{3.2}. The complete
energy--momentum is defined by 
\be
T_\pm =T_\pm^\one +T_\pm^\two \, ,                           \lab{4.4} 
\ee
and it also satisfies the Virasoro algebra \eq{3.2}.

The Poisson bracket algebra between $I_{\pm a}$ and $T_\pm$ has the
form   
\bea
&&\{I_{\pm a}(\s_1),I_{\pm b}(\s_2)\}
                            =f_{ab}{^c}I_{\mp c}(\s_2)\d\, ,\nn\\
&&\{T_\pm(\s_1),I_{\pm a}(\s_2)\}=-I_{\pm a}(\s_1)\d'\, , \nn\\
&&\{T_\pm(\s_1),T_\pm(\s_2)\}=-[T_\pm(\s_1)+T_\pm(\s_2)]\d'\, ,
                                                             \lab{4.5}
\eea
and represents two copies of the semi--direct product of the $SL(2,R)$
and Virasoro algebras. Together with diffeomorphisms, described in the
previous section, we have here an additional $SL(2,R)\times SL(2,R)$
structure. Therefore, the collection $(T_\pm, I_{\pm a})$ can be taken
as a set of {\it first class constraints\/} in the general canonical
construction based on \eq{1.3}.  The related dynamical system will be
called the {\it WZNW system\/}. 

We display here the complete set of constraints, multipliers and 
gauge parameters:
$$ 
\ba{lllll}  
G_m=   & T_-,   & T_+,   & I_{-a},   & I_{+a},  \\
u^m=   & h^-,   & h^+,   & a^a_+,    & a^a_-,   \\
\ve^m= & \ve^-, & \ve^+, & \eta^a_+, & \eta^a_-.
\ea
$$

Now, using 
\bsubeq 
\lab{4.6}
\be
H_0=0\, ,\qquad G_m=(T_-,T_+,I_{-a},I_{+a}) \, ,            \lab{4.6a}
\ee
one can construct the related canonical Lagrangian:  
\be
\cL(q_i,\pi_i,h)=\pi_{1\a}q^\a_1+\pi_{2\a}q^\a_2 -h^-T_- -h^+T_+ 
                   -a^a_+I_{-a}-a^a_-I_{+a} \, .            \lab{4.6b}
\ee
\esubeq 
It represents a gauge theory invariant under both local 
$SL(2,R)\times SL(2,R)$ transformations and diffeomorphisms.
   
Using the general rule \eq{1.4}, with $V_r{^m}=0$ and $U_{sr}{^m}$
calculated from the algebra \eq{4.5} (Appendix C), one finds the
following gauge transformations of the multipliers:
\bea
&&\d h^\pm=(\pd_0 +h^\mp\pd_1)\ve^\pm-\ve^\pm\pd_1 h^\pm \, ,\nn\\
&&\d a^c_\pm=(\pd_0+h^\mp\pd_1)\eta^c_\pm -f_{ab}{^c}a^a_\pm\eta^b_\pm
                              -\ve^\mp\pd_1 a^c_\pm \, .     \lab{4.7}
\eea
Gauge transformations of the dynamical variables are 
\be
\d q^\a = -\bE^\a{_a}\eta^a_+ -E^\a{_a}\eta^a_- 
            +{1\over 2\k}\bigl( \ve^+ J^\a_+ -\ve^-J^\a_-\bigr) \, ,
                                 \qquad q=q_1 \, ,            \lab{4.8}
\ee
while $\d q_2$ is obtained by changing $\k$ to $-\k$.

As before, we can eliminate the momenta $\pi_{1\a}$ and $\pi_{2\a}$
in order to clarify the usual Lagrangian content of the theory.
Explicit calculation shows the complete agreement with the Lagrangian
treatment of reference \cite{12}: the resulting Lagrangian describes
the gauge extension of the WZNW system \eq{1.1}, invariant under local
$SL(2,R)\times SL(2,R)$ and diffeomorphisms (Appendix D).  

In the canonical approach, the WZNW system is introduced in the process
of constructing the first class constraints $I_{\pm a}$, used to gauge
the whole $SL(2,R)\times SL(2,R)$ group.  
This approach closely parallels the related gauge procedure in the
Lagrangian formalism \cite{12}. Namely, it is well known that one can 
not gauge the simple WZNW model \eq{2.1} consistently for an arbitrary
gauge group $H\subseteq SL(2,R)\times SL(2,R)$, since the Wess--Zumino
term $\G$ does not have a gauge invariant extension that can be
expressed as an integral over spacetime $\S$ \cite{16}. However, the
problem can be solved by going over to the WZNW system \eq{1.1}, where
the problematic, nonlocal term appearing in the first sector during the
gauge procedure, cancels the corresponding term in the second sector,
producing thus the consistent gauge theory for every $H$.  
There is a clear analogy between the cancellation of nonlocal terms in
the Lagrangian procedure, and the elimination of central charges in the
Hamiltonian approach.  

\section{Gauge extension of the WZNW system and \protect\\
         induced gravity} 
\cleq

In this section we shall show that 2D induced gravity can be obtained
from the canonical gauge extension of the WZNW system, by 
\bitem
\item[(a)] performing a suitable gauge fixing, and 
\item[(b)] integrating out some dynamical variables in the functional
integral.  
\eitem

\subsection{Canonical \mb{H_+\times H_-} gauge theory} 
\cleq

Let us consider a restriction of the canonical theory \eq{4.6},
defined by the following subset of first class constraints: 
\be
G'_m\;=\;\bigl( T_-,\;T_+,\; I_n\bigr)\, ,\qquad  
I_n\equiv\bigl[I_{-\gp},I_{-\gn},I_{+\gm},I_{+\gn}\bigr]\, ,  \lab{5.1}
\ee
representing a subalgebra of \eq{4.5}. This restriction can be obtained
from the full canonical theory \eq{4.6} by imposing the following
gauge conditions: 
\be
a^\gm_+=0 \, ,\qquad a^\gp_-=0 \, .                           \lab{5.2}
\ee
The restricted algebra based on \eq{5.1} describes diffeomorphisms
combined with the internal symmetry
\be
H=H_+\times H_- \, ,                                           \lab{5.3}
\ee
where $H_+$ and $H_-$ are subgroups of $SL(2,R)$ defined by the
generators $(t_+,t_0)$ and $(t_0,t_-)$, respectively.

The canonical action of the restricted theory takes the form  
\be
\cL(q_i,\pi_i,h) =\pi_{1\a}\dot q^\a_1 + \pi_{2\a}\dot q^\a_2 
                   - h^-T_- - h^+T_+ -a^n I_n \, ,             \lab{5.4}
\ee
where $a^n \equiv [ a^\gp_+,a^\gn_+,a^\gm_-,a^\gn_-]$.
Here, explicit expressions for the energy--momentum components are
given by equation \eq{4.4}, in conjunction with \eq{4.3} and \eq{3.1},
while the currents $I_n$ are of the form
\bea
&&I_{-\gp}= \pi_{x_1}+\pi_{x_2} \, ,\nn \\
&&I_{-\gn}= \bigl[ x_1\pi_{x_1} +(\pi_{\vphi_1}-\k\vphi'_1)\bigr]
        +\bigl[ x_2\pi_{x_2}+(\pi_{\vphi_2}+\k\vphi'_2)\bigr]\, ,\nn\\
&&I_{+\gm}= -\pi_{y_1}-\pi_{y_2} \, ,\nn\\
&&I_{+\gn}= \bigl[ -y_1\pi_{y_1} -(\pi_{\vphi_1}+\k\vphi'_1)\bigr]
        +\bigl[ -y_2\pi_{y_2} -(\pi_{\vphi_2}-\k\vphi'_2)\bigr]\, .\nn
\eea

It is clear that the canonical action \eq{5.4} represents a gauge
extension of the WZNW system \eq{1.1}. Indeed,  by choosing the gauge
fixing $a^n=0$, and eliminating the momenta $\pi_{1\a}$ and
$\pi_{2\a}$, the action \eq{5.4} reduces to the form  
$$
\cL(q_1,q_2,h)=\cL(q_1,h)-\cL(q_2,h) \, ,
$$
where $\cL(q,h)$ is given by equation \eq{3.6}, representing the
covariant extension of \eq{1.1} (see Appendix D). In what follows we
shall demonstrate that the action \eq{5.4} can be effectively reduced
to the induced gravity \eq{1.2}. 

\subsection{Effective theory in the canonical form} 

\subsub{Quantum action.} In order to demonstrate the connection of
the canonical theory \eq{5.4} to the induced gravity, we begin by
choosing the gauge conditions corresponding to the first class
constraints $I_n$:    
\bea
&&\O_n\equiv\bigl[\O_{-\gp},\O_{-\gn},\O_{+\gm},\O_{+\gn}\bigr]\, ,\nn\\
&&\O_{\mp(\pm)} =J^\one_{\mp(\pm)}-\m_{\mp} =0\, ,\qquad
  \O_{\mp\gn} =J^\two_{\mp\gn}-\l_{\mp} =0\, .                 \lab{5.5}
\eea 
To impose these conditions in the functional integral, we use
the BRST formalism and introduce the following set of ghosts,
antighosts and new multipliers: 
$$  
\ba{llll} 
\hbox{\rm Ghost~fields:}    & e^-, & e^+, & c^n,     \\
\hbox{\rm Antighosts:}      &      &      & \bc^n,   \\
\hbox{\rm Multipliers:}     &      &        & b^n,    
\ea
$$
where $c^n\equiv [c^{-\gp}, c^{-\gn}, c^{+\gm}, c^{+\gn}]$, and
similarly for $\bc^n$ and $b^n$. 
While ghost fields correspond to gauge parameters, antighosts and
multipliers are associated to the gauge conditions. Since the
diffeomorphisms are not gauge fixed, the related antighosts and
multipliers are not present in the formalism. 
The BRST transformation $sX$ of a dynamical variable $X$,
$X=(q_1,q_2,h^\pm)$, is obtained from the gauge transformation 
$\d X$ by replacing gauge parameters with ghosts; for the new fields 
we have $s\bc^n=b^n, sb^n=0$, while $sc^n$ is not needed here 
($sc^n$ follows from the nilpotency condition: $s^2X=0$). 

Then, we introduce the gauge fermion $\Psi= \bc^n\O_n$, and define the
quantum action in the usual way: 
\be
\cL_Q =\cL(q_i,\pi_i,h)+s\Psi
      =\cL(q_i,\pi_i,h) + \cL_{GF} +\cL_{FP} \, .            \lab{5.6}
\ee
The gauge fixing and the Faddeev--Popov parts are given by
$$ 
\cL_{GF}=b^n\O_n \, ,\qquad \cL_{FP}=-\bc^n [s\O_n]\, ,
$$
where 
\bea
&&s\O_{\mp(\pm)} = -\bigl[ e^\mp J_{\mp(\pm)}^\one\bigr]' 
                            \mp c^{\mp\gn}J_{\mp(\pm)}^\one \, ,\nn\\
&&s\O_{\mp\gn} = -\bigl[e^\mp J_{\mp\gn}\bigr]' \pm
  c^{\mp(\pm)}J_{\mp(\pm)}^\two \mp 2\k\bigl[ c^{\mp\gn}\bigr]'\, .\nn
\eea

\subsub{Effective theory.} Having derived the quantum action, we are
now going to show that it can be effectively reduced to the induced
gravity, by integrating out all the variables except $\vphi_1,\vphi_2$,
and the related momenta. To simplify the exposition technically, we
shall divide it into several smaller steps. 

(a) The integration over the multipliers $b^\pm$, $a_+$ and $a_-$
transforms $\cL_Q$ into the effective Lagrangian   
$$
\cL_E(\vphi_i,\pi_{\vphi_i},h)
  =\bigl[ \pi_{1\a}\dot q_1^\a+\pi_{2\a}\dot q_2^\a
                  -h^-T_- -h^+T_+ +\cL_{FP}\bigr]_{I=\O=0} \, .         
$$
It is now convenient to rewrite the first class constraints 
$I_n=0$ and the related gauge conditions $\O_n=0$ in the form
$$
J^\one_{\mp(\pm)}=\m_\mp = -J^\two_{\mp(\pm)} \, ,\qquad
-J^\one_{(\mp)\gn}=\l_\mp = J^\two_{\mp\gn} \, ,      
$$
or, more explicitly,
\bea
&&\pi_{x_1}=\m_-=-\pi_{x_2}\, ,\nn\\
&&-\pi_{y_1}= \m_+ =\pi_{y_2} \, ,\nn\\
&&x_1\pi_{x_1}+2K_{1-}   =\l_-=-(x_2\pi_{x_2}+2K_{2+}) \, ,\nn\\
&&-(y_1\pi_{y_1}+2K_{1+})=\l_+= y_2\pi_{y_2}+2K_{2-} \, , \nn
\eea
where $K_\pm=(\pi_\vphi \pm \k\vphi')/2$. 

(b) The momentum variables $\pi_{x_1},\pi_{y_1}$ and
$\pi_{x_2},\pi_{y_2}$ are constant, so that the related $\pi\dot q$
terms in the action can be ignored as total time derivatives. 

(c) Also, the contribution of the Faddeev--Popov term is decoupled since
the currents $J^\one$ and $J^\two$ are constant, so that the
integration over ghosts and antighosts can be absorbed into the
normalization of the functional integral. 

(d) Finally, the expression for $T_\mp$, reduced to the surface $I=\O=0$, 
reads
$$
\k\tT_\mp =\bigl[\pm (K_{1\mp})^2 +2\k(K_{1\mp})'\bigr] +
          \bigl[\mp (K_{2\pm})^2 +2\k(K_{2\pm})'\bigr]
          \mp \fr{1}{4}\m\bigl( e^{\vphi_1} -e^{\vphi_2}\bigr)\, , 
$$
where $\m=\m_-\m_+$, so that the effective theory in the canonical form
is given by   
\be
\cL_E(\vphi_i,\pi_{\vphi_i},h)
  =\pi_{\vphi_1}\dot\vphi_1 +\pi_{\vphi_2}\dot\vphi_2 
                              -h^-\tT_- -h^+\tT_+\, .         \lab{5.7}
\ee

\subsection{Transition to the induced gravity} 

In order to find out the usual dynamical content of the previous
result, we shall eliminate the remaining momentum variables from \eq{5.7}
by using their equations of motion:
\bea
&&\pi_{\vphi_1}={\k\over\sqrt{2}}\bigl[ \hpd_-\vphi_1 +\hpd_+\vphi_1 
                           +2(\hcon_- -\hcon_+)\bigr] \, ,\nn\\
&&\pi_{\vphi_1}\pm\k\vphi_1 =\sqrt{2}\k\bigl( \hpd_\pm\vphi_1 
                           +\hcon_- -\hcon_+ \bigr) \, ,      \lab{5.8}
\eea
while  $\pi_{\vphi_2}$ is obtained by the replacement
$\vphi_1\to\vphi_2, \k\to -\k$ ($\hpd_\pm$ and $\ho_\pm$ are defined in
Appendix B). The effective theory is described by the Lagrangian
\bea
&&\cL_E(\vphi_1,\vphi_2,h)=\L(\vphi_1,h)-\L(\vphi_2,h) \, ,\nn\\
&&\L(\vphi,h)=\sqrt{-\hg}\bigl[ \k\hpd_+\vphi\hpd_-\vphi
              +2\k \bigl(\hcon_-\hpd_+\vphi -\hcon_+\hpd_-\vphi\bigr) 
              +Me^\vphi\bigr] \, ,                            \lab{5.9}
\eea
where $M=\m/2\k$. If we now change the variables according to
$$
\phi=\sqrt{\k}(\vphi_1-\vphi_2) \, ,\qquad 2F=\vphi_2 \, ,
$$
the effective Lagrangian takes the final form:
\bea
\cL_E(\phi,F,h)=\sqrt{-\hg}\Bigl\{ \hpd_+\phi\hpd_-\phi
      && +2\sqrt{\k} \bigl[ (\hcon_- +\hpd_-F)\hpd_+\phi 
                        -(\hcon_+ -\hpd_+F)\hpd_-\phi \bigr]  \nn\\
      && +Me^{2F}\bigl( e^{\phi/\sqrt{\k}} -1\bigr)\Bigr\} \, . \nn
\eea

The geometric meaning of this Lagrangian becomes more transparent if
we use conformally rescaled metric (Appendix B),
$$
g_{\m\n}=e^{2F}\hg_{\m\n}\, ,\qquad 
         \con_\pm =e^{-F}\bigl(\hcon_\pm \mp \hpd_\pm F\bigr) \, ,\qquad
         \pd_\pm =e^{-F}\hpd_\pm \, ,
$$
whereupon the effective Lagrangian is easily seen to transform into the
expression that coincides with the induced gravity action \eq{1.2}:
\bea
\cL_E(\phi,g_{\m\n})&&=\sqrt{-g}\bigl[ \pd_+\phi\pd_-\phi
      +2\sqrt{\k} \bigl(\con_-\pd_+\phi -\con_+\pd_-\phi\bigr) 
      +M\bigl( e^{\phi/\sqrt{\k}} -1\bigr) \bigr]       \nn\\
                  &&=\sqrt{-g}\bigl[ \pd_+\phi\pd_-\phi
      +\sqrt{\k}\phi R +M\bigl( e^{\phi/\sqrt{\k}} -1\bigr)\bigr] \nn\\
                  &&=\cL_G(\phi, g_{\m\n})  \, ,             \lab{5.10}
\eea
where we used equation \eq{B8}.

\section{Geometric properties from gauge transformations} 
\cleq

In the process of constructing the induced gravity action from the
gauged WZNW system, one expects the original {\it gauge\/}
transformations of dynamical variables to go over into 
{\it geometric\/} transformations of the final, gravitational theory.
We have already seen that gauge transformations of canonical
multipliers $h^\pm$ produce correct geometric transformations 
of the metric density $\tg^{\m\n}$. Complete interpretation of the
induced gravity demands to clarify the nature of two additional fields,
$\sqrt{-g}$ and $\phi$, given by
\be
\sqrt{-g}=\fr{1}{2}(h^- -h^+)e^{\vphi_2} \, , \qquad 
\phi=\sqrt{\k}(\vphi_1-\vphi_2) \, .                      \lab{6.1}
\ee
 
We begin by noting that the transformation rule \eq{4.8} of the WZNW
variables $q_i=(x_i,\vphi_i,y_i)$, $i=1,2$, describes the $SL(2,R)$
gauge transformations, defined by parameters $\eta_\pm$ \cite{12}, and 
the $\ve^\pm$ transformations, which we expect to be related to
diffeomorphisms. In particular, the $\ve^\pm$ transformation of
$\vphi_1$ has the form  
\be
\d_\ve \vphi = -{1\over 2\k}\bigl[ \ve^+(\pi_\vphi +\k\vphi') 
               -\ve^-(\pi_\vphi -\k\vphi'\bigr] \, ,\qquad 
                                         \vphi=\vphi_1 \, ,  \lab{6.2}
\ee
while $\d_\ve\vphi_2$ is obtained by replacing $\k\to -\k$.

Now, let us go to the gauge fixed, effective theory, expressed by
equation \eq{5.7}. While the gauge transformations in the WZNW theory
are defined using the Poisson brackets in \eq{1.4}, the related
transformation rules in the gauge fixed theory (induced gravity) should
be calculated with the help of the {\it Dirac brackets\/}, determined
by $(I_n,\O_n)$.  

In order to check whether $\d(\sqrt{-g})$ has the correct geometric
form \eq{B5}, we use the results of Appendix E, in particular equation
\eq{E3}, and find that the above transformation law for $\vphi$ should
be replaced with the Dirac bracket expression:
\be
\d^*_\ve\vphi =\d_\ve\vphi -\pd_1(\ve^- +\ve^+) \, .         \lab{6.3}
\ee
where, after eliminating $\pi_\vphi$ with the help of \eq{5.8}, 
$\d_\ve\vphi$ takes the form  
\bea
\d_\ve\vphi &&={1\over\sqrt{2}}\bigl[ 
                  -(\ve^+\hpd_+\vphi -\ve^-\hpd_-\vphi) +
                   (\ve^- -\ve^+)(\hcon_- -\hcon_+)  \bigr]  \nn\\
            &&=-\ve\cdot\pd\vphi -\ve^0\pd_1(h^- +h^+) \, .  \nn
\eea
Comparing the expression \eq{6.3} with equation \eq{B6}, one
concludes that $\d^*_\ve\vphi$ yields the correct transformation law
for $\sqrt{-g}$.   

It is now easy to see that the variable $\phi$ behaves as a scalar
field,  
\be
\d_\ve^*\phi =-\ve\cdot\pd\phi \, ,                         \lab{6.4} 
\ee
in agreement with its geometric role. 

The following relations make the geometric structure of the effective
theory particularly transparent: 
\bea
&&\{ T^\one_\pm(\s_1),T^\one_\pm(\s_2)\}^*= 
   -[T^\one_\pm(\s_1)+T^\one_\pm(\s_2)]\d \mp 2\k \d''' \, ,\nn\\
&&\{ T^\two_\pm(\s_1),T^\two_\pm(\s_2)\}^*= 
   -[T^\two_\pm(\s_1)+T^\two_\pm(\s_2)]\d \pm 2\k \d''' \, ,\nn\\ 
&&\{ T_\pm(\s_1),T_\pm(\s_2)\}^*= -[T_\pm(\s_1)+T_\pm(\s_2)]\d \, .
                                                            \lab{6.5}
\eea
We see that the energy--momentum components of the WZNW sectors 1 
and 2, $T^\one$ and $T^\two$ are not first class constraints in
the gauge fixed, effective theory, since their algebras contain
central charges, while the complete energy--momentum tensor,
$T=T^\one+T^\two$, is of the first class.   

\section{Concluding remarks} 
\cleq

In the present paper we used the canonical approach to elucidate how
the induced gravity action, together with its geometric properties, can
be obtained from the dynamical structure of the $SL(2,R)$ WZNW system. 

We first analyzed chiral symmetries of the $SL(2,R)$ WZNW model \eq{2.3},
using the Hamiltonian formalism based on the choice of time
$\t=\xi^\pm$, which led us naturally to the currents $J_{\pm a}$,
satisfying two independent $SL(2,R)$ KM algebras. These currents are
basic objects in our canonical approach. They are used to construct
quadratic $SL(2,R)$ invariants, the energy--momentum components 
$T_\pm$, that satisfy two independent Virasoro algebras and represent
first class constraints corresponding to diffeomorphisms, in the
canonical gauge formalism defined by \eq{1.3}. Then, the gauge
procedure is generalized by introducing two sets of KM currents,
$J^\one_{\pm a}$ and $J^\two_{\pm a}$, corresponding to two sectors of
the WZNW system \eq{1.1}, which are used to define the new first class 
constraints $I_{\pm a}=J^\one_{\pm a}+J^\two_{\pm a}$, satisfying an
$SL(2,R)\times SL(2,R)$ algebra without central charge, and the
energy--momentum components $T_\pm$ corresponding to the whole WZNW
system. The resulting theory is clearly gauge equivalent to the 
WZNW system \eq{1.1}, being its canonical gauge extension. 
The Hamiltonian process of elimination of central charges in
the algebra of new currents $I_{\pm a}$ closely parallels the
cancellation of the nonlocal terms in the Lagrangian approach
\cite{12}. As the main result of our analysis, we showed, (a) by
choosing a suitable gauge fixing, and (b) integrating out some
dynamical variables, that this gauge theory reduces effectively to the
induced gravity \eq{1.2}. Geometric properties of the gravitational
theory are derived from gauge properties of the gauge extended WZNW
system, with the help of the Dirac brackets.  

The results obtained here supplement those of the recent Lagrangian
analysis \cite{12}, and improve our understanding of geometric
properties of 2D spacetime in terms of the related gauge structure.
They can be used to better understand singular solutions of the induced
gravity in terms of globally regular solutions of the WZNW system, and
clarify the nature of black holes \cite{9,17}.     

\section*{Acknowledgment} 

This work was supported in part by the Serbian Science Foundation,
Yugoslavia. 

\appendix 

\section{Geometric properties of \mb{SL(2,R)}} 

In this Appendix we present some useful results concerning the
Riemannian structure of the group manifold $SL(2,R)$.  

Choosing the generators of $SL(2,R)$ as
$t_{(\pm)}=\fr{1}{2}(\s_1\pm i\s_2)$, $t_\gn =\fr{1}{2}\s_3$, where
$\s_k$ are the Pauli matrices, one finds that the Lie algebra
$[t_a,t_b]=f_{ab}{^c}t_c$ takes the form  
\be
[t_\gp,t_\gm]=2 t_\gn \, ,\qquad [t_{(\pm)},t_\gn]=\mp t_{(\pm)}\, .
                                                             \lab{A1}
\ee
Explicit evaluation of the Cartan metric 
$\g_{ab}=(t_a,t_b)={\fr 1 2}f_{ac}{^d}f_{bd}{^c}$ yields 
\be
\g_{ab}=\pmatrix{ 0 &0 &2\cr
                  0 &1 &0\cr
                  2 &0 &0\cr} \, ,\qquad a,b={\gp,\gn,\gm}\, .\lab{A2}
\ee
The Cartan metric $\g_{ab}$ and its inverse $\g^{ab}$ are used to lower
and raise the tangent space indices $(a,b,...)$. 

Any element $g$ of $SL(2,R)$ in a neighborhood of identity can be
parametrized by using the Gauss decomposition:  
\be
g=e^{xt_\gp}e^{\vphi t_\gn}e^{yt_\gm}
 = e^{-\vphi/2}\pmatrix{ e^\vphi +xy & x \cr
                             y       & 1 \cr }\, ,            \lab{A3}
\ee 
where $q^\a=(x,\vphi,y)$ are group coordinates. 

Now, the Lie algebra valued 1--form 
$v=g^{-1}dg=t_aE^a=t_a E^a{_\a}dq^\a$ defines the quantity $E^a{_\a}$,
the vielbein on the group manifold. The above expression for $g$ leads
to  
\bea
&&E^\gp=e^{-\vphi}dx \, ,\nn \\
&&E^\gn=2ye^{-\vphi}dx +d\vphi \, ,\nn \\
&&E^\gm=-y^2e^{-\vphi}dx -yd\vphi +dy \, , \nn
\eea
so that the vielbein $E^a{_\a}$ and its inverse $E^\a{_a}$ are given as
\be
E^a{_\a} = \pmatrix{   e^{-\vphi} &    0&   0\cr
                      2ye^{-\vphi}&    1&   0\cr
                    -y^2e^{-\vphi}&   -y&   1\cr} \, ,\qquad
E^\a{_a} = \pmatrix{   e^\vphi &    0&     0\cr
                        -2y    &    1&     0\cr
                        -y^2   &    y&     1\cr} \, .         \lab{A4}
\ee
The Cartan metric in the coordinate basis, 
$\g_{\a\b}=E^a{_\a}E^b{_\b}\g_{ab}$, has the form 
\be
\g_{\a\b}=\pmatrix{    0      &  0& 2e^{-\vphi}\cr
                       0      &  1&      0\cr 
                   2e^{-\vphi}&  0&      0\cr }  \, ,
                                 \qquad \a,\b=x,\vphi,y\, .  \lab{A5}
\ee

Using the property $d(v,v^2)=0$, one can write locally $(v,v^2)=-6d\t$,
where  
\be
d\t=E^\gp E^\gn E^\gm =d\bigl( e^{-\vphi}dx dy \bigr)\, .    \lab{A6}
\ee

Similarly, the calculation of $\bv=gdg^{-1}=t_a\bE^a=t_a\bE^a{_\a}dq^\a$ 
leads to
\be
\bE^a{_\a}=\pmatrix{-1 &    x &   x^2e^{-\vphi}\cr
                        0 &   -1 &   -2xe^{-\vphi}\cr
                        0 &    0 &   -e^{-\vphi}  \cr} \, ,\qquad
\bE^\a{_a}=\pmatrix{ -1 &   -x &      x^2\cr
                      0 &   -1 &      2x \cr
                      0 &    0 &   -e^{\vphi}  \cr} \, .      \lab{A7}
\ee
The metric $\bar\g_{\a\b}$ is the same as $\g_{\a\b}$.

\section{Riemannian structure on \mb{\S}} 
\cleq

Here, we present some basic features of the Riemannian geometry on  
two--dimensional spacetime $\S$.

\subsub{Light--cone basis.} Starting from the interval on $\S$,
$$
ds^2=g_{\m\n}d\xi^\m d\xi^\n 
    =(d\xi^0)^2\bigl[g_{00}+2g_{01}u+g_{11}u^2\bigr]\, ,
     \qquad u\equiv {d\xi^1/d\xi^0} \, ,
$$
we can solve the equation $ds^2=0$ for $u$,
$$
u_{1,2}={-g_{01}\pm \sqrt{-g}\over g_{11} }\equiv h^\pm \, , 
$$
and obtain
$$
ds^2=(d\xi^0)^2g_{11}(u-h^+)(u-h^-) = 2d\xi^+d\xi^-\, .
$$
Here,
\bea
&&d\xi^+=\sqrt{-g_{11}/2}\,(-h^+d\xi^0 +d\xi^1)=e^+{_\m}d\xi^\m\, ,\nn\\
&&d\xi^-=\sqrt{-g_{11}/2}\,(h^-d\xi^0-d\xi^1)=e^-{_\m}d\xi^\m \, .\nn
\eea
If we introduce $-g_{11}=e^{2F}$, three independent components of the
metric $g_{\m\n}$ can be expressed in terms of the new, light--cone
variables $(h^-,h^+,F)$. In particular, 
$$
\sqrt{-g}=e^{2F}\sqrt{-\hg}\, ,\qquad 
                      \sqrt{-\hg}\equiv \fr{1}{2}(h^- -h^+)\, .
$$
At each point of $\S$ the quantities  
\be
e^i{_\m}=e^F\,\he^i{_\m} \, ,\qquad
\he^i{_\m}\equiv {1\over\sqrt{2}}\pmatrix{ -h^+ &1 \cr
                                            h^- &-1\cr }\qquad (i=+,-)
                                                             \lab{B1}
\ee
define an orthonormal, light--cone basis of 1--forms,
$\th^i=d\xi^i=e^i{_\m}d\xi^\m$. Note that 
$\det(e^i{_\m}) =-\sqrt{-g}$. We also introduce the 
related basis of tangent vectors, $e_i\equiv\pd_i=e_i{^\m}\pd_\m$, 
\be
e_i{^\m}=e^{-F}\,\he_i{^\m}\, ,\qquad
 \he_i{^\m}\equiv {\sqrt{2}\over h^- -h^+}\pmatrix{ 1   &h^-   \cr
                                                    1   &h^+ \cr}\, .
                                                            \lab{B2}
\ee
The metric $\eta_{ij}$ in the tangent space has the light--cone form:
$\eta_{-+}=\eta_{+-}=1$. Tangent space components of an arbitrary
vector $V_\m$ are
\be
V_\pm =e_\pm{^\m}V_\m =e^{-F}\,{\sqrt{2}\over h^- -h^+}(V_0+h^\mp V_1)
                      =e^{-F}\,{\hV}_\pm \, .               \lab{B3}
\ee 
The components of the metric $\hg_{\m\n}$ and its inverse $\hg^{\m\n}$
are 
\bea
&&\hg_{\m\n}= {1\over 2}\pmatrix{ -2h^-h^+& h^-+h^+ \cr
                                  h^-+h^+&   -2     \cr } \, ,\nn \\
&&\hg^{\m\n}= {2\over (h^--h^+)^2}\pmatrix{  2    & h^-+h^+\cr
                                          h^-+h^+&  2h^-h^+\cr }\, .\nn
\eea

\subsub{Diffeomorphisms.} The standard transformation rule of the
zweibein $e^i{_\m}$ under the diffeomorphisms, 
$\xi^\m\to\xi^\m+\ve^\m(\xi)$, implies  
\be
\d h^\pm=\pd_0\ve^\pm +h^\pm\pd_1\ve^\pm -\ve^\pm\pd_1 h^\pm\, .
                                                            \lab{B4} 
\ee
where $\ve^\pm=\ve^1-\ve^0 h^\pm$. 
The transformation rule of the metric is: 
\bea
&&\d g^{\m\n}=g^{\m\r}\pd_\r\ve^\n +g^{\n\r}\pd_\r\ve^\m 
                                   -\ve^\r\pd_\r g^{\m\n} \, ,\nn \\
&&\d \sqrt{-g}= -\pd_\r(\ve^\r\sqrt{-g}) \, .               \lab{B5}
\eea
The conformal rescaling $g_{\m\n}=e^{2F}\hg_{\m\n}$ leads to 
$$
\d F=-\pd_1\ve^1 +\pd_1\ve^0 (h^- +h^+) -\ve^\m\pd_\m F \, .
$$
Transition to $\ve^\pm$ yields
\be
\d(2F)=-\pd_1(\ve^+ +\ve^-) 
         +(\ve^- -\ve^+){\pd_1(h^- +h^+)\over h^- -h^+}
         -{1\over\sqrt{2}}(\ve^+\hpd_+ -\ve^-\hpd_-)2F \, .  \lab{B6}
\ee 

The algebra of diffeomorphisms has the form
$$
[\d(\ve_{1}),\d(\ve_{2})]h^\pm =\d(\ve_{3})h^\pm \, ,
    \qquad \ve^\pm_{3}=\ve^\pm_{1}\partial_1\ve^\pm_{2}
                              -\ve^\pm_{2}\partial_1\ve^\pm_{1} \, .
$$
It is similar to the Virasoro algebra, but not the same, since 
$\ve^\pm=\ve^\pm(\xi^+,\xi^-)$. 
In the limit of conformally flat space, $h^\pm \to \mp 1$,
coordinate transformations are restricted to two sets of conformal
transformations, $\ve^\pm=\ve^\pm(\xi^\pm)$, with two independent
Virasoro algebras. The light--cone gauge is defined by $h^+=-1$ 
(or $h^-=1)$ and $\sqrt{-g}=1$.    

\subsub{Connection and curvature.} Riemannian connection on $\S$ is 
defined by the first structural equation: 
$$
d\th^i +\con^i{_j}\wedge\th^j =0 \, ,\qquad \con^i{_j}=\ve^i{_j}\o\, .
$$ 
For the connection 1--form $\con=\con_i\th^i$ we find
\bea
&&\con_+=e^{-F}(\hcon_+ -\hpd_+F)\, ,\qquad
                  \hcon_+=-{\sqrt{2}\over h^- -h^+}(h^-)' \, ,\nn\\
&&\con_-=e^{-F}(\hcon_- +\hpd_+F)\, ,\qquad
                  \hcon_-={\sqrt{2}\over h^- -h^+}(h^+)' \, . \lab{B7}
\eea

The curvature is defined by the second structural equation:
$$
d\con^i{_j}=\fr{1}{2}R^i{_{jkl}}\,\th^k\wedge\th^l\, , 
$$
where we used $\con^i{_k}\wedge\con^k{_j}=0$. Since
$d\con=(\nabla_-\con_+-\nabla_+\con_-)\th^-\wedge\th^+$, one finds
\be
R=2R_{+-}=2(\nabla_-\con_+ -\nabla_+\con_-) \, .            \lab{B8}
\ee

\section{Structure functions} 
\cleq

The Poisson bracket algebra \eq{4.5} implies the relations
\bea
&& \{ (\eta^{\mp a}I_{\mp a})_\s, (a^b_\pm I_{\mp b})_{\s_2}\} 
      = [f_{ab}{^c}\eta^{\mp a}a^b_\pm]_\s (I_{\mp c})_\s \d\, ,\nn\\
&& \{ (\eta^{\mp a}I_{\mp a})_\s, (h^\mp T_\mp)_{\s_2} \}
      = [(\eta^{\mp a})'h^\mp]_\s (I_{\mp a})_\s \d \, ,\nn \\
&& \{ (\ve^\mp T_\mp)_\s,(a^a_\pm I_{\mp a})_{\s_2} \}
      = - [\ve^\mp(a^a_\pm)']_\s(I_{\mp a})_\s \d \, ,\nn \\
&& \{ (\ve^\mp T_\mp)_\s, (h^\mp T_\mp)_{\s_2}\} 
  = [-(h^\mp)'\ve^\mp +(\ve^\mp)'h^\mp]_\s (T_\mp)_\s\d \, ,\nn
\eea 
needed to calculate gauge transformations of the multipliers
$u^m=(h^-,h^+,a^a_+,a^a_-)$, according to the general rule \eq{1.4}. 
Here, $\d=\d(\s-\s_2)$, and one understands that an integration over
$\s$ and $\s_2$ is to be performed. 

\section{Lagrangian form of the gauged WZNW system} 
\cleq

In this Appendix we find the usual Lagrangian description of the
canonical gauge action for the WZNW system.

First, we focus our attention on the restricted action \eq{5.4}, which
can be written as a sum of two terms, $\cL=\cL_1+\cL_2$, describing
two sectors of the theory. The variation over $\pi_{x_1}$,
$\pi_{\vphi_1}$ and $\pi_{y_1}$ yields:    
\bea
&&\pi_{x_1}=2\sqrt{2}\k e^{-\vphi_1}\hD_-y_1 \, ,\nn \\
&&\pi_{\vphi_1} \pm \k\vphi_1'=\sqrt{2}\k 
          \bigl[\hpd_\pm\vphi_1 +\hA^\gn_+ -\hB^\gn_- \bigr]\, ,\nn \\
&&\pi_{y_1}=2\sqrt{2}\k e^{-\vphi_1} \hD_+ x_1  \, ,\nn
\eea
where
\bea
&&\hA^c_+ = {-\sqrt{2}\over h^- -h^+}\, a^c_+ \, ,\qquad
  \hB^c_- = {-\sqrt{2}\over h^- -h^+}\, a^c_- \, ,\nn \\
&&\hD_+x_1=\bigl[ \hpd_+ +\hA^\gn_+\bigr]x_1 +\hA^\gp_+  \, ,\qquad
  \hD_-y_1=\bigl[\hpd_- -\hB^\gn_-\bigr]y_1 -\hB^\gm_- \, .  \lab{D1}
\eea
Replacing this into $\cL_1$ leads to
\bea
&&\cL_1=\cL^r(q_1,\hA,\hB) 
        +\k\sqrt{-\hg} \bigl[ \hB^\gn_- -\hA^\gn_+\bigr]^2 \, ,\nn \\
&&\cL^r(q_1,\hA,\hB)=\k\sqrt{-\hg}\bigl[\hpd_+\vphi_1\hpd_-\vphi_1
            + 2\hA^\gn_+\hpd_-\vphi_1 -2\hB^\gn_-\hpd_+\vphi_1
                   +4\hD_+ x_1\hD_- y_1 e^{-\vphi_1} \bigr] \, .\nn
\eea

In a  similar manner one finds
$$
\cL_2 =-\cL^r(q_2,\hA,\hB) 
        -\k\sqrt{-\hg}\bigl[ \hB^\gn_- -\hA^\gn_+\bigr]^2 \, ,
$$
so that the complete Lagrangian of the gauged WZNW system \eq{5.4} is
given by 
\be
\cL=\cL_1+\cL_2=\cL^r(q_1,\hA,\hB)-\cL^r(q_2,\hA,\hB) \, .   \lab{D2}
\ee

Comparing this result with equation (4.5) in Ref. \cite{12}, one finds
that \eq{D2} gives the standard Lagrangian description of the gauged 
WZNW system, provided the canonical multipliers $\hA_+$ and $\hB_-$ are
identified with the related gauge potentials. The correctness of this
identification is checked by comparing their transformation laws.
Thus, e.g., $\eta_\pm$ gauge transformations of the canonical
multipliers, obtained with the help of \eq{4.7}, 
$$
\d_\eta\hA^c_+ =-\hpd_+\eta^c_+ -f_{ab}{^c}\hA^a_+\eta^b_+\, ,\qquad
\d_\eta\hB^c_- =-\hpd_-\eta^c_- -f_{ab}{^c}\hB^a_-\eta^b_-\, , 
$$
are identical to the $SL(2,R)\times SL(2,R)$ transformations of
Lagrangian gauge potentials. 

Similar analysis can be done for the complete gauge action \eq{4.6b},
with the same conclusion.

\section{Dirac brackets} 
\cleq

In order to have a clear geometric interpretation of the transition
from the gauged WZNW system to the induced gravity, we shall calculate
here the Dirac brackets corresponding the set of first class
constraints and gauge conditions:   
\bea
&&\th_{+a}=\bigl(I_{+\gm},\O_{+\gn},I_{+\gn},\O_{+\gm}\bigr) \, ,\nn\\
&&\th_{-a}= \bigl(I_{-\gp}, \O_{-\gn},I_{-\gn},\O_{-\gp}\bigr) \, , 
                                                            \lab{E1}
\eea 
The calculation of $\D_{+ab}=\{\th_{+a},\th_{+b}\}$ yields:
$$
(\D_+)_{ab}=\pmatrix{      0&  -\m_+\d&        0&        0& \cr 
                      \m_+\d&   2\k\d'&   2\k\d'&        0& \cr
                           0&   2\k\d'&        0&  -\m_+\d& \cr 
                           0&        0&   \m_+\d&        0& \cr} \, .
$$
The inverse is:
$$
(\D_+)^{-1ab}={1\over\m_+^2}
             \pmatrix{ 2\k\d'&   \m_+\d&         0&  -2\k\d'& \cr 
                      -\m_+\d&        0&         0&        0& \cr
                            0&        0&         0&   \m_+\d& \cr 
                      -2\k\d'&        0&   -\m_+\d&        0& \cr} \, .
$$
Similarly,
$$
(\D_-)_{ab}= -(\D_+)_{ab}\vert_{\m_+\to\m_-} \, ,\qquad
(\D_-)^{-1ab}= -(\D_+)^{-1ab}\vert_{\m_+\to\m_-} \, .
$$
Since $\{\th_{+a},\th_{-a}\}=0$, the Dirac bracket of $X$ and $Y$ is
defined by 
\be
\{ X,Y\}^*=\{ X,Y\} -\{ X,\th_{+a}\} (\D_+)^{-1ab} \{\th_{+b},Y\}
           -\{ X,\th_{-a}\} (\D_-)^{-1ab} \{\th_{-b},Y\} \, . \lab{E2}
\ee

We display here several useful results:
\bea
&&\{\vphi_1,\pi_{\vphi_1}\}^*=\{\vphi_1,\pi_{\vphi_1}\}\, , \qquad
  \{\vphi_1, T^\one_\pm\}^*=\{\vphi_1, T^\one_\pm\} -\d' \, ,\nn \\
&&\{\vphi_2,\pi_{\vphi_2}\}^*=\{\vphi_2,\pi_{\vphi_2}\}\, , \qquad
\{\vphi_2, T^\two_\pm\}^*=\{\vphi_2, T^\two_\pm\} -\d' \, .    \lab{E3}
\eea


\end{document}